\documentclass{article}
\usepackage{arxiv}
\usepackage[utf8]{inputenc} 
\usepackage[T1]{fontenc}    
\usepackage{hyperref}       
\usepackage{url}            
\usepackage{booktabs}       
\usepackage{amsfonts}       
\usepackage{graphicx} 
\usepackage{orcidlink}
\usepackage{placeins}
\usepackage{longtable}
\usepackage{pdflscape}
\usepackage{titlesec}
\usepackage{biblatex}
\title{Doctoral Theses in France (1985-2025): A Linked Dataset of PhDs, Academic Networks, and Institutions}
\author{\orcidlinki{William Aboucaya\thanks{These authors contributed equally. Name order is alphabetical.}}{0000-0003-2413-6968} \\
	ACSS -- DRM\\
	Université Paris Dauphine -- PSL\\
	Paris, France \\
	\texttt{william.aboucaya@dauphine.psl.eu} \\
	\And
	\orcidlinki{Dastan Jasim\footnotemark[1]}{0000-0002-4158-276X} \\
	ACSS -- DRM\\
	Université Paris Dauphine -- PSL\\
	Paris, France \
	\texttt{dastan.jasim@dauphine.psl.eu} \\
}
\date{}

\renewcommand{\shorttitle}{A Linked Dataset of PhDs, Academic Networks, and Institutions}

\begin{document}
\maketitle

\setcounter{footnote}{0}

\begin{abstract}
This paper presents a comprehensive dataset of doctoral theses defended in France between 1985 and 2025, constructed from multiple national academic metadata sources. The dataset is primarily based on data from the French national thesis platform and is enriched using additional authority and bibliographic databases to improve data quality, completeness, and interoperability. The data production pipeline includes the aggregation of heterogeneous sources, the correction of inconsistent identifiers, the enrichment of person and institution records, and the construction of derived variables describing academic careers, jury participation, institutional affiliations, and thesis characteristics. Additional identifiers from major academic repositories and library catalogues are integrated to facilitate linkage with external data sources and future dataset extensions. The resulting dataset provides structured information at the thesis, individual, and institutional levels, enabling both descriptive and relational analyses. This resource is particularly suited for research on doctoral education, academic networks, supervision practices, jury composition, institutional collaboration, and the evolution of research communities over time. The paper documents the data sources, processing pipeline, feature construction, data quality issues, and limitations, with the objective of facilitating reuse of the dataset by other researchers and supporting future extensions and longitudinal analyses of the academic system.
\end{abstract}

\section{Introduction}

Academic networks and the trajectories of the scholars embedded in them have attracted sustained empirical attention~\cite{makes_corsini_2022,comment_godechot_2010,historical_huang_2019, gender_lerchenmueller_2018, phd_villarroya_2008}. However, few of these have systematically used data on PhD juries to do so. Doctoral theses and their defenses, however, constitute a rich resource for the study of such academic systems. Each defense brings together a candidate whose supervisors are, in most cases, people they have worked with over many years, and who, in turn, invite jury members who are often linked to the directors and the candidates for multiple reasons. Disciplinary, institutional, and sometimes even personal relations come into play here, enabling rich longitudinal analyses of supervision practices, jury composition, co-occurrence of researchers, disciplinary evolution, and even career trajectories. Yet few national datasets make such information available in a form amenable to systematic quantitative research. While bibliographic metadata sources exist and provide information about PhD thesis authors, titles, and other details, few integrate interoperable identifiers like the \textit{Thèses.fr} dataset. Furthermore, the dataset provides information on Ph.D. juries and identifiers for jury members as well\footnote{https://www.data.gouv.fr/datasets/theses-soutenues-en-france-depuis-1985}.

This data paper presents an enriched and corrected dataset derived from \textit{Thèses.fr}, a platform maintained by the ABES (Agence bibliographique de l'enseignement supérieur -- Bibliographic Agency for Higher Education) covering all doctoral theses defended in France between since 1985. Starting from the raw metadata that is publicly available, we apply a multi-stage pipeline that enriches the data using the academic identifier IdRef and optimizes it, for example, by using an inference tool to impute missing values for actors' genders based on their names. Variables measuring the centrality of academic actors in co-occurrence networks of PhD juries are also provided to help users make longitudinal analyses of academic networks. The resulting dataset is further linked to complementary national infrastructures such as TEL and SUDOC, thereby enhancing interoperability with full-text repositories and bibliographic catalogues via these identifiers.

While the authors' use case focuses on gendered networks, the dataset can address many questions at the intersection of scientometrics, sociology of science, and network analysis, including structures of supervision, career trajectories, and topical developments. It also covers processes of internationalization, as the dataset includes non-French PhD candidates who completed their doctorates in France. The paper first provides an overview of related work and existing research gaps, then details the dataset's production. It then describes the dataset and offers guidance on its reuse.

We give access to the dataset through an open data repository\footnote{\url{https://doi.org/10.5281/zenodo.19453191}} and code used to produce it is available from GitHub\footnote{\url{https://github.com/WilliamAboucaya/phd-theses-france/issues}}.

\section{Related works}

Different researchers and institutions have undertaken projects to standardize the archiving, aggregation, and dissemination of metadata on doctoral theses in Europe and beyond. While scraping and enriching the Theses.fr dataset, we systematically surveyed comparable infrastructures. Across the examined systems, a central distinction emerged: whether a platform operates as (i) an administrative register aligned with the doctoral defense process and therefore more likely to contain defense dates, locations, and structured committee roles, or (ii) a bibliographic union catalogue or harvesting aggregator optimized for discoverability and interoperability, typically exposing standardized bibliographic fields and identifiers but not modeling the defense event itself.

Over the past two decades, several initiatives sought to harmonize and centralize thesis discovery across Europe. The DART-Europe e-theses portal\footnote{\url{https://www.ucl.ac.uk/library/dart-europe-e-theses-portal-has-closed-down}} operated as a pan-European harvesting service that aggregated metadata for open-access theses from institutional repositories via OAI-PMH, providing a centralized discovery layer linking to locally hosted full texts. The portal was permanently closed on 3 February 2025. Its discontinuation reflects a broader shift from specialized thesis portals toward general open-access aggregators such as BASE\footnote{\url{https://www.base-search.net}}, CORE\footnote{\url{https://core.ac.uk}}, and NDLTD’s Global ETD Search\footnote{\url{https://ndltd.org/resources/find-etds/}}, which index theses alongside other scholarly outputs. Although these services enable large-scale bibliographic aggregation and improved searchability, they do not structure doctoral defense events or committee roles as queryable entities. This trajectory illustrates both the effort to standardize thesis discovery at the supranational level and the subsequent deconsolidation of dedicated infrastructures.

At the national level, many systems follow a similar aggregation logic. A first natural comparison for our study was Germany. We contacted the German National Library (Deutsche Nationalbibliothek - DNB) in March 2025 to inquire about obtaining a full dataset of defended theses with identifiers comparable to Theses.fr. A fully open, bulk-downloadable dataset does not exist. Dissertations can be retrieved via the DNB portal using the search term \texttt{hss=diss*}\footnote{\url{https://portal.dnb.de/opac/simpleSearch?query=hss\%3Ddiss*&cqlMode=true}} and filtered by "Hochschulschriften" (university theses), with export available through the data shop in XML, JSON, or CSV formats. However, author-level metadata such as date of birth or gender are not systematically provided. ORCID identifiers appear only when recorded in corresponding GND authority entries, limiting consistent linkage to external person-level data. As shown for the United States~\cite{porter2025understanding}, ORCID uptake varies substantially across disciplines, and similar heterogeneity is observable in Germany. Crucially, DNB records do not include structured information on defense events, such as committee membership, roles, or defense location. For research focused on the gendered composition and co-occurrence structures of doctoral juries, these variables are unavailable, although the dataset remains valuable for bibliographic and authority-based analyses.

Comparable aggregation-based infrastructures exist elsewhere in Europe. In Sweden, DiVA\footnote{\url{https://www.diva-portal.org}} aggregates theses and other scholarly outputs from multiple universities, offering harmonized bibliographic metadata and persistent identifiers. In Portugal, RCAAP\footnote{\url{https://www.rcaap.pt}} harvests institutional repositories to provide national-level access to theses and dissertations. While supervisor or jury information may appear in local records or full texts, committee composition is not consistently normalized or exposed as standardized metadata at the aggregation layer. In the Czech Republic, Theses.cz\footnote{\url{https://theses.cz}} centralizes thesis records and full texts nationally but likewise prioritizes bibliographic description over structured modelling of defense committees. Outside Europe, ProQuest Dissertations \& Theses\footnote{\url{https://about.proquest.com/en/products-services/pqdtglobal/}} offers extensive global coverage through subscription-based access and standardized bibliographic fields, yet defense-committee data are not exposed in reusable structured form.

A structurally closer comparator to Theses.fr is Teseo, which collects information on theses defended in Spain since 1976, including supervisors and committee members\footnote{\url{https://aplicaciones.ciencia.gob.es/teseo/\#/home}}. Ricardo González-Haba scraped this database for network analysis and released the resulting dataset via the Open Science Framework~\cite{gonzalezhaba2023teseo}. He constructed weighted inter-university networks based on shared participation in doctoral thesis defenses between 1977 and 2022, identifying geographically coherent communities and increasing intra-university concentration of ties. However, the released dataset lacks systematic gender or authority identifiers, limiting demographic enrichment. Earlier, Manuel Blázquez-Ochando compiled a downloadable version of the Teseo catalogue by programmatically crawling the Ministry’s website and structuring thesis-level metadata including title, author, university, defense date, supervisors, and committee members into a relational database~\cite{blazquez2015tesis}\footnote{\url{https://mblazquez.es/catalogo-de-tesis-doctorales-espanolas-teseo-disponible-para-su-descarga/}}. While enabling bibliometric and network-oriented analyses, this work did not include systematic authority control or demographic enrichment.

In contrast to aggregation-based infrastructures and prior scraping efforts, our enrichment of Theses.fr extends beyond bulk extraction. We validated and corrected authority identifiers such as IdRef, enriched person- and institution-level metadata from external authority files, inferred missing demographic attributes including gender, and engineered longitudinal and role-based indicators. By leveraging the fact that Theses.fr is embedded in the French administrative doctoral governance process and structurally models defense events and jury roles, our dataset enables reconstruction of committee configurations and co-occurrence networks at scale.

What further distinguishes Theses.fr and Teseo from harvested datasets is their integration into a regulated administrative data production and transmission process. Rather than collecting metadata after publication from heterogeneous repositories, both systems generate thesis records within the formal doctoral workflow. In Spain, doctoral programs are regulated under the "Real Decreto 99/2011, de 28 de enero", and the TESEO database is administered by the Spanish Ministry of Science, Innovation, and Universities. Universities supply validated thesis information, including defense date, institution, supervisors, and tribunal members, as part of this reporting process, after which records are published nationally\footnote{\url{https://www.ciencia.gob.es/Universidades/TESEO.html}}. This integration ensures that the defense event is structurally modelled rather than incidentally described.

Despite this shared administrative logic, important differences remain. Theses.fr is integrated into the French national bibliographic infrastructure through the French Bibliographic Agency for Higher Education’s regulated thesis reporting network, which applies standardized cataloguing rules and links persons and institutions to persistent authority identifiers such as IdRef\footnote{\url{https://abes.fr/reseau-theses/le-reseau/}}. Teseo, although likewise a ministerial register of defended theses, relies on centrally reported administrative data without systematic integration of persistent person identifiers or a comparable national authority control system. Both systems, therefore, enable reconstruction of defense networks, but Theses.fr provides a higher degree of authority normalization and interoperability, whereas Teseo requires more extensive cleaning and enrichment to achieve comparable analytical robustness.

\section{Production of the dataset}

This section describes the end-to-end pipeline used to construct the theses dataset analyzed in this work. The production process builds upon the aggregation of multiple authoritative bibliographic and authority sources and proceeds through successive stages of integration, enrichment, and correction. Starting from raw metadata retrieved from existing national platforms, the pipeline combines heterogeneous data sources to produce a unified and structured dataset suitable for large-scale analysis. In addition to data collection, the process includes the engineering of new features, the normalization and alignment of entities, and the identification and correction of inconsistencies present in the original sources. These corrections are performed using a combination of automated methods and targeted human interventions, with the objective of improving data quality, completeness, and reliability while preserving traceability to the original records.

\subsection{Sources}

The dataset is primarily derived from a main source, \textit{Thèses.fr}, the official French national platform for the dissemination of doctoral theses, maintained by the \textit{Agence Bibliographique de l’Enseignement Supérieur} (Higher Education Bibliographic Agency, ABES). Thèses.fr provides comprehensive coverage of doctoral research in France by indexing all Ph.D. theses defended since 1985, as well as doctoral theses currently in preparation when they have been formally reported by academic institutions. In addition to core thesis records, the platform provides structured information on the individuals involved in each thesis like doctoral candidates, supervisors, and jury members, as well as on the associated institutions. Each record typically includes rich bibliographic metadata such as titles, abstracts, keywords, languages, and disciplinary classifications -- designed to facilitate search, retrieval, and downstream analysis. Data from Thèses.fr are made openly available through a public API\footnote{\url{https://www.data.gouv.fr/dataservices/api-export-des-donnees-de-theses-fr}} and can also be downloaded as bulk datasets in CSV, JSON, or NDJSON formats\footnote{\url{https://www.data.gouv.fr/datasets/theses-soutenues-en-france-depuis-1985}}. The data are made available under the Etalab Open License, the open-data license adopted by most French public institutions. 
To construct the dataset, we relied on the JSON bulk export, as this format provides more complete data than what can be retrieved through API queries or other export formats. However, this export was published on January 8, 2024. Consequently, all theses defended after this date, as well as theses that had already been defended but had not yet been reported to ABES at the time of the dataset publication, were not included in the bulk export. To mitigate this limitation and improve temporal coverage, we complemented the dataset by retrieving missing thesis records for the years 2022 to 2025 using the public API. This complementary data collection step was performed on March 31, 2026. As a result, the dataset used in this study reflects the state of the \textit{Thèses.fr} database as of that date. Users of the dataset should therefore be aware that theses reported to ABES after this date are not included and that future updates of the dataset may further improve coverage for the most recent years.

To complement and enrich this information, we additionally rely on \textit{IdRef} (Identifiants et Référentiels), the French national authority database for higher education and research, also maintained by ABES. IdRef provides persistent identifiers and authority records for entities involved in scholarly production, including persons (e.g., researchers, authors, supervisors), organizations (e.g., universities and research institutions), and conferences. These authority records consolidate variant names, affiliations, and contextual information with the goal of enabling entity disambiguation and ensuring consistency across bibliographic data sources. Thèses.fr includes IdRef identifiers for most individuals and institutions referenced in thesis records, which allows for reliable linkage and cross-referencing between theses, persons, and organizations. IdRef data are openly accessible via a public API, downloadable in structured formats such as RDF and JSON, and can also be queried through a SPARQL endpoint\footnote{\url{https://data.idref.fr/sparql}}, enabling advanced semantic queries and data integration. Similar to Thèses.fr, IdRef data are also distributed under the Etalab Open License.

\subsection{Data enhancements}

After downloading the raw data from the source platforms, we apply a series of data enhancement steps aimed at improving overall data quality prior to feature engineering and analysis. These enhancements focus on correcting inconsistencies and omissions in the original sources that would otherwise limit interoperability, enrichment, or downstream analytical validity.

A first enhancement step concerns the validation and correction of IdRef identifiers provided in the Thèses.fr dataset. We observe that certain IdRef references are ill-formed or do not correspond to any existing IdRef authority record, preventing reliable linkage with external authority data. These invalid identifiers were individually reviewed and corrected by one of the authors of this paper, based on contextual information available in the thesis records and cross-referencing with the IdRef database. After this step, 168 IdRef identifiers have been corrected and refer to the correct record. This manual correction step enables the subsequent enrichment of person- and institution-level metadata using authoritative identifiers.

A second enhancement step addresses the incomplete coverage of gender information in IdRef authority records for persons. As gender is not systematically recorded in IdRef, many individual records lack this attribute. To mitigate this limitation, we apply an automated gender inference tool that assigns a probable gender based on the first name when sufficient confidence can be established. We have chosen to use the \texttt{nomquamgender}~\cite{Van_Buskirk_Clauset_Larremore_2023} Python package which proposes state-of-the-art level performances for gender inference from first name and has already been evaluated in the context of the Spanish doctoral theses platform TESEO~\cite{Matias-Rayme2024}. 
This process allowed us to improve the completion of our gender data from 433,340 persons with unknown gender to 37,621 after gender guessing.

After those correction steps, we enrich the initial Thèses.fr dataset with the data provided by IdRef on the persons and institutions involved in theses. First, we found that in many instances, the first and last names of persons were switched in the Thèses.fr. Therefore, we decided to keep the names from the IdRef instead as they are typically more reliable. Additionally, for each person in a thesis record, we enriched their data with the following information from IdRef when they were available:

\begin{itemize}
    \item Gender
    \item Birth date
    \item Death date
    \item Language(s) of writing
    \item Country
\end{itemize}

Once person-level records had been enriched and validated, we proceeded to construct additional derived features designed to facilitate quantitative and longitudinal analyses. These features were computed at both the individual and thesis levels and primarily aim to capture aspects of academic seniority and participation intensity that are not directly available in the raw sources.

At the individual level, we compute age at the time of the thesis defense for all persons involved in a given thesis, provided that their date of birth is available in the corresponding IdRef authority record. This feature enables comparative analyses of academic trajectories and generational patterns across roles (e.g., doctoral candidates, supervisors, jury members).

For Ph.D. supervisors and jury members specifically, we also derive additional role-related indicators intended to approximate academic experience and involvement in doctoral evaluation processes:

\begin{itemize}
    \item The number of occurrences in juries during the four years preceding the thesis defense, computed dynamically with respect to each defense date. This variable captures recent participation intensity in doctoral evaluation. The four-year time window is motivated by the typical duration of doctoral studies in France, which generally is between 36 and 52 months~\cite{Le_Chanu_2025}. Restricting the measure to this period ensures that the indicator reflects evaluation activity occurring within a timeframe directly relevant to the preparation and supervision of the thesis under consideration.
    \item The number of years since the individual defended their own Ph.D. thesis, when this information is available, used as an approximation for academic seniority.
    \item The number of years since the individual first participated in a doctoral jury (either as a jury member or as a supervisor), providing an estimate of experience in doctoral supervision and evaluation.
\end{itemize}

Finally, we introduce thesis-level features describing linguistic and organizational characteristics of each defense. We identify whether the thesis was primarily written in French or in another language, based on the declared language metadata. We also compute the number of distinct languages in which a title or abstract is provided, which serves as an indicator of the intended international dissemination scope. In addition, we also include the total number of Ph.D. supervisors, jury members, reviewers, and distinct academic institutions associated with each thesis. 

Together, these engineered features substantially extend the analytical potential of the dataset beyond its original descriptive metadata, enabling more refined investigations of academic careers and institutional dynamics in doctoral education.

\subsection{Interoperability}

To further facilitate the reuse of our dataset and enable its extension through additional external data sources, we also integrate two complementary identifiers for each thesis record. These identifiers correspond to those used by \textit{Thèses En Ligne} (Online Theses, TEL) and the \textit{Système Universitaire de Documentation} (University Documentation System, SUDOC), two major infrastructures providing metadata for theses and academic documents in France. By including these identifiers, we allow our dataset to be easily linked with other repositories that contain complementary bibliographic, institutional, or full-text information.

TEL is the thesis-specific subplatform of the French open-access academic archive HAL. Within this repository, Ph.D. graduates can voluntarily deposit the full manuscript of their thesis together with associated metadata, such as the authors' academic affiliations, keywords, and subject classifications drawn from a structured taxonomy of research topics\footnote{Available at \url{https://theses.hal.science/browse/domain}}. However, because thesis deposition in TEL is performed voluntarily by doctoral graduates or their institutions, the corresponding TEL identifier is only available for a subset of theses. 

In contrast, SUDOC is a national catalogue that aggregates bibliographic records from academic libraries and documentation centers across France. One of its missions is to systematically catalogue doctoral theses produced in French institutions. For each thesis manuscript, SUDOC records detailed bibliographic metadata, including the document format (e.g., printed volume, microfiche, or digital version) and the libraries or documentation centers where the thesis can be consulted. Each record is assigned a unique identifier called the \emph{Pica Production Number} (PPN). Because theses are systematically catalogued by libraries, this identifier is available for the vast majority of theses in our dataset.

The inclusion of both TEL and SUDOC identifiers significantly enhances the interoperability of the dataset by allowing direct linkage with additional metadata sources and full-text repositories. As a result, they enable researchers to derive new features or enrich existing records using information maintained in these external platforms.

\section{Description of the dataset}

\begin{figure}
    \centering
    \includegraphics[width=\linewidth]{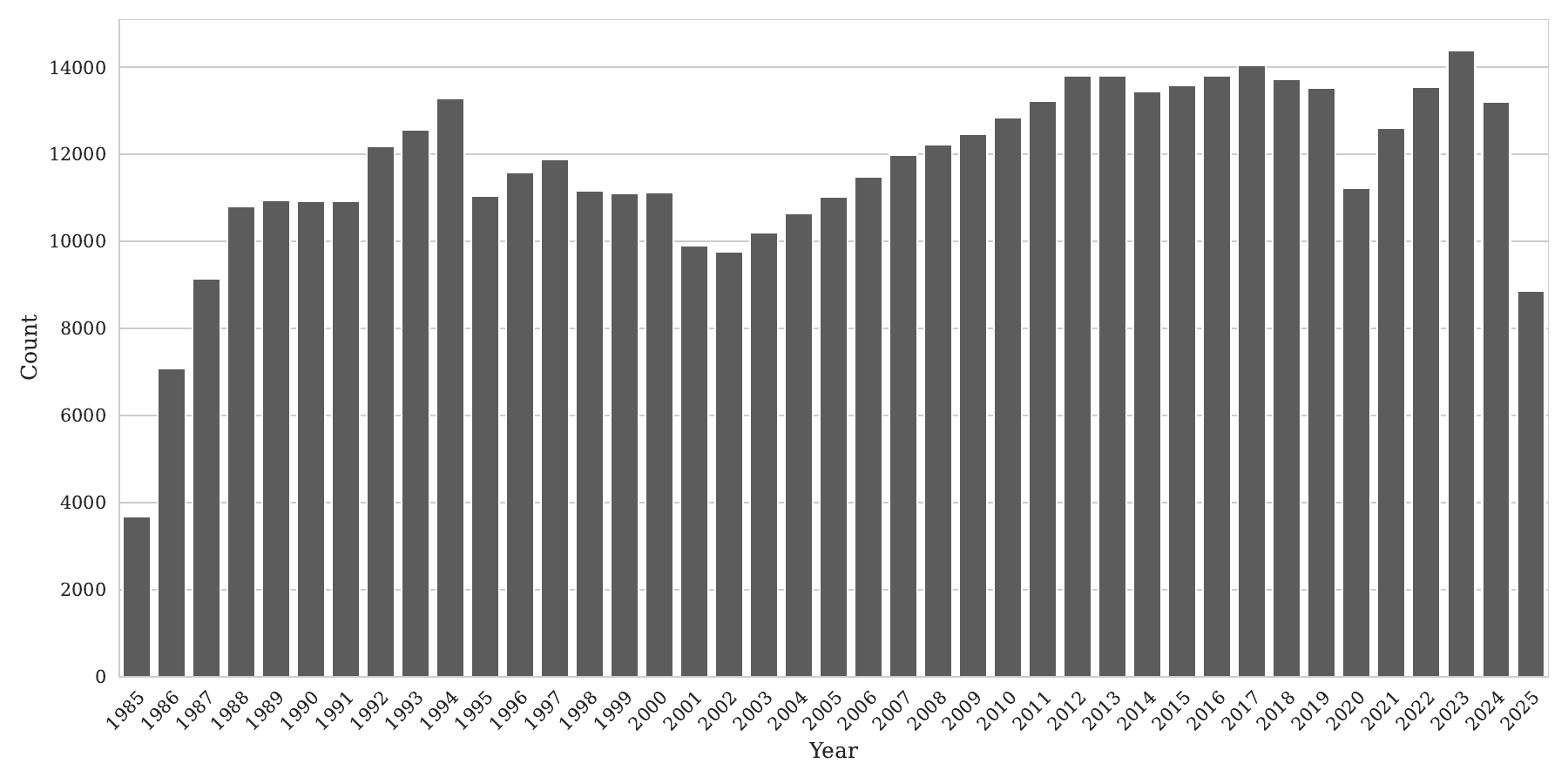}
    \caption{Number of theses in the dataset per year}
    \label{fig:theses per year}
\end{figure}

\begin{figure}
    \centering
    \includegraphics[width=\linewidth]{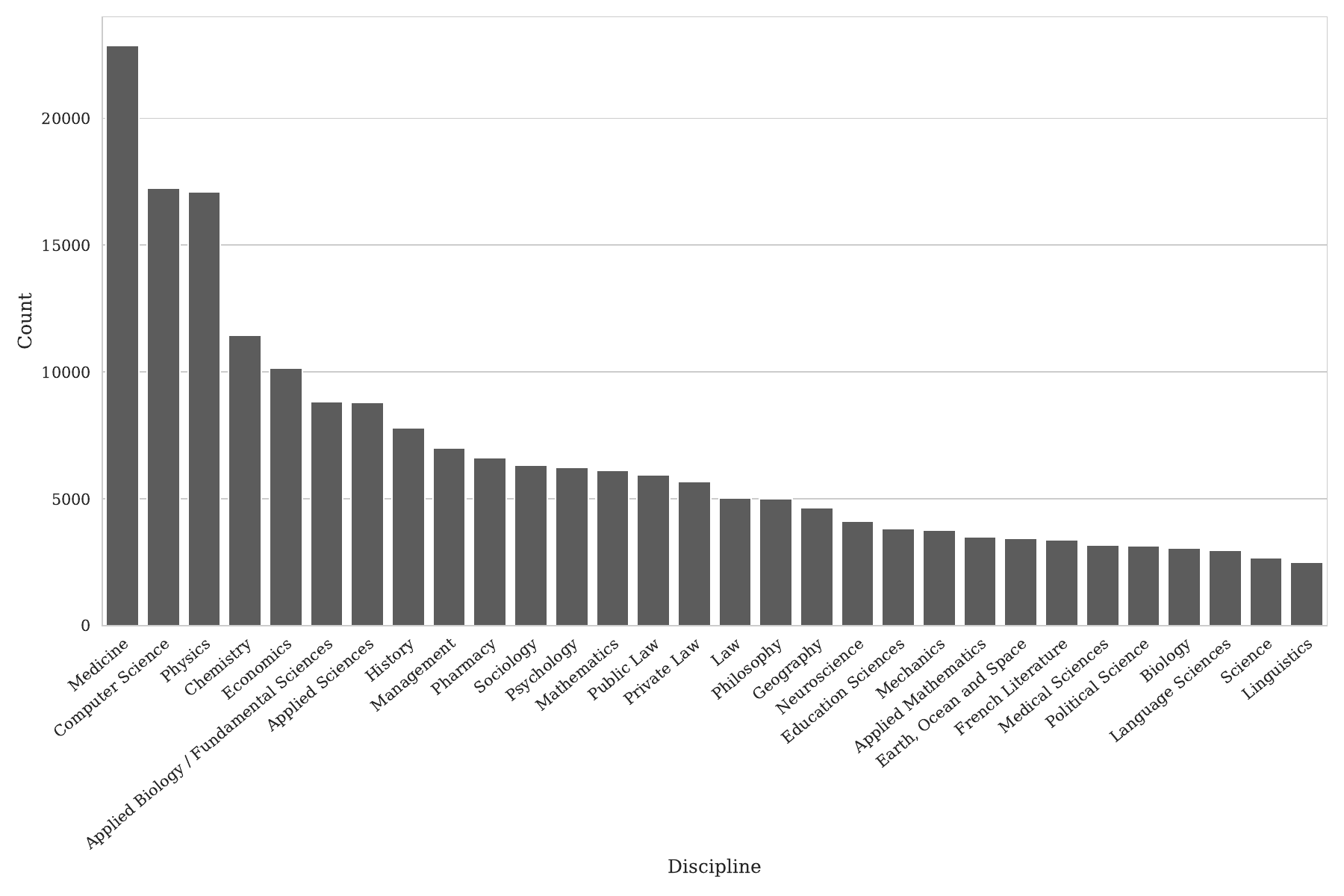}
    \caption{Distribution of research topics (translated from French)}
    \label{fig:discipline distribution}
\end{figure}

Each row of the dataset corresponds to a single Ph.D. thesis that has been defended in France between 1985 and 2025 (see Figure~\ref{fig:theses per year} for year-based distribution). A wide variety of research fields are covered, such as medicine, sociology, theology or art studies (see Figure~\ref{fig:discipline distribution}). The dataset is structured so as to capture not only bibliographic information about the thesis itself, but also detailed information about the individuals involved in the defense, their institutional affiliations, and the content and thematic classification of the thesis. For clarity and usability, the variables included in the dataset are organized into six main categories, described below.

\begin{itemize}
    \item \textbf{Thesis identifiers \& status:} This category includes unique identifiers associated with the thesis across different bibliographic infrastructures, as well as administrative and status-related information such as the type of Ph.D., the defense date, and publication status of the manuscript.

    \item \textbf{Author:} This category contains information describing the doctoral candidate, including personal identifiers and demographic or biographical information when available.

    \item \textbf{Supervisors:} This category includes information about the thesis supervisor or co-supervisors, including identifiers, demographic information, and derived indicators related to academic experience and jury participation.

    \item \textbf{Jury:} This category contains information about all members of the thesis defense committee, including rapporteurs, jury members, and the jury president.

    \item \textbf{Institutional affiliations:} This category describes the research institution(s) involved in the thesis defense, the doctoral school(s) and the research partner(s) associated with the thesis.

    \item \textbf{Content \& topics:} This category includes textual and thematic information describing the thesis. It contains the title, abstract, and free-text keywords in multiple languages, as well as predefined topical classifications used to categorize the thesis by discipline and research area.
\end{itemize}

An exhaustive list of variables is provided in Table~\ref{tab:variables} of Appendix~\ref{appendix:variables codebook}. In addition, Figures~\ref{fig:theses_missing} to \ref{fig:subject_entries_missing} of Appendix~\ref{appendix:missing data} present the proportion of missing values for selected variables in the dataset to document data completeness and potential limitations for empirical analyses. Reporting missing data is particularly crucial in this dataset as the availability of certain features varies significantly across time periods and across thesis records. In particular, Figure~\ref{fig:dir_jury_missing} shows that a large proportion of theses do not contain detailed information about the composition of the defense jury. This is primarily due to historical data collection practices: when the \textit{Thèses.fr} platform was first introduced, information about jury members was not systematically included in thesis records. As a result, older theses are less likely to contain detailed jury information, while more recent records tend to be more complete. This temporal heterogeneity in data completeness should therefore be taken into account when performing longitudinal analyses involving jury composition or participation.

\begin{figure}
    \centering
    \includegraphics[width=\linewidth]{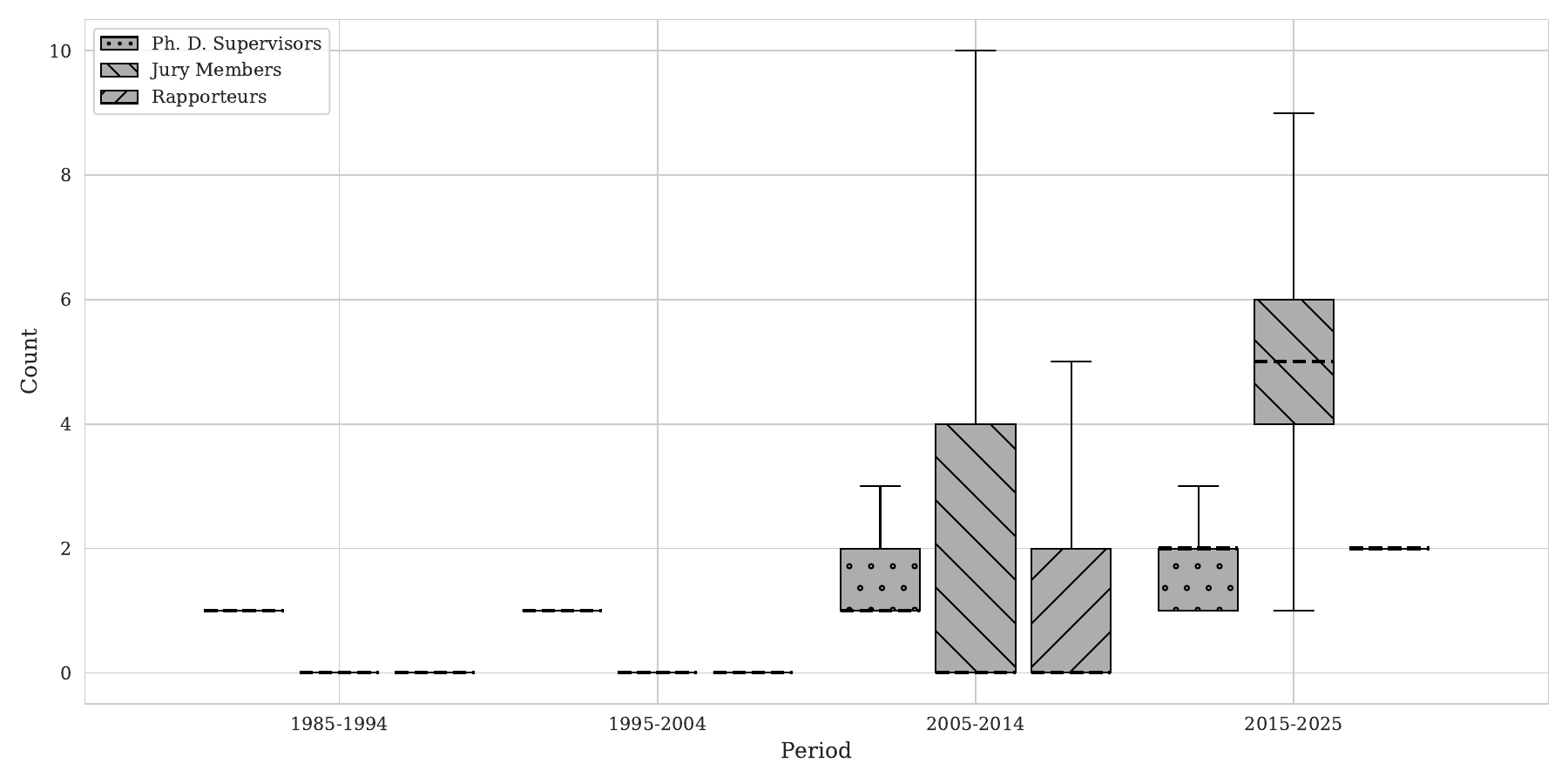}
    \caption{Distribution of the number of Ph. D. supervisors, jury members and thesis rapporteurs for each decade.}
    \label{fig:distribution persons per period}
\end{figure}

Figure~\ref{fig:distribution persons per period} further shows that the average number of supervisors, jury members, and thesis rapporteurs recorded per thesis increases over time. This trend can be explained by several factors. First, metadata recording practices have improved over time, leading to more systematic inclusion of jury-related information in thesis records. Second, jointly supervised theses have become more common, contributing to a higher number of supervisors per thesis. Third, changes in doctoral regulations regarding thesis jury composition have encouraged larger and more diverse juries~\cite{Legifrance2016cadredoctorat}.

The case of doctoral schools is similar. Doctoral schools are organizational entities within universities that oversee for example doctoral training, administrative procedures, or aspects of doctoral supervision and evaluation. However, these structures were not systematically established in French universities before being formally institutionalized at the national level by regulatory reforms implemented in 2006~\cite{Legifrance2006ecoledoctorale}. Prior to this reform, experimental structures performing similar functions existed, but only in some universities or regions. As a result, information related to doctoral schools is almost systematically absent from thesis records before 1990 and remains very sparse between 1990 and 2006. This absence should therefore not be interpreted as missing data in the usual sense, but rather as the consequence of the historical evolution of doctoral education institutions in France. After 2006, the presence of doctoral school information in thesis records becomes much more systematic, reflecting the institutionalization of doctoral schools.

Certain features, for example those describing a thesis jury member, are made available multiple times because they describe an entity which can be present more than once for a given thesis. These versions are differentiated by a numerical index appended to the variable name (e.g., \texttt{title.0}, \texttt{jury\_member.1}, etc.), allowing access to information about each individual version of the entity. For each indexed feature, the index values are inherited directly from the original \textit{Thèses.fr} dataset. To the best of our knowledge, this order is arbitrary and does not reflect an explicit hierarchy in the thesis record. For example, \texttt{language.0} is not necessarily the primary language of the thesis manuscript. For each group of columns associated with a person (author, supervisor, rapporteur, jury member, or jury president), the dataset includes their IdRef identifier, first name, last name, gender (when available or inferred), country, birth and death dates, as well as information related to their languages of writing (when available).

Another feature naming pattern concerns the variables describing the textual content of the thesis itself, namely its title, abstract, and free-form topic keywords. Because these textual elements may be available in multiple languages for a single thesis, multiple versions of each feature are included in the dataset. For each of these indexed feature, an additional column with the same base name and the suffix \texttt{.language} is provided (e.g., \texttt{title.0.language}). This companion variable specifies the language associated with the corresponding textual field, allowing users to identify and filter content by language. This structure makes it possible to preserve all available multilingual information while maintaining a tabular dataset format. Although most theses in the dataset are written primarily in French ($\sim 88\%$), a large proportion of theses also include an English version of their titles ($\sim 81\%$) and abstracts ($\sim 59\%$). The distribution of languages across all textual variables is presented in Table~\ref{tab:language_dist}. 

\begin{landscape}
\begin{table}
\caption{Language distribution across thesis metadata fields. Column headers show \% of theses with no language info recorded.}
\label{tab:language_dist}
\footnotesize
\begin{tabular}{llllllll}
\toprule
Language & Thesis
(0.0\% NA) & Abstract
(12.8\% NA) & Title
(0.0\% NA) & Subject
(41.7\% NA) & Author
(14.0\% NA) & Supervisor
(3.7\% NA) & Jury
(61.3\% NA) \\
\midrule
French & 414554 (86.601\%) & 416536 (87.015\%) & 471600 (98.518\%) & 278841 (58.25\%) & 357350 (74.651\%) & 454423 (94.93\%) & 184341 (38.509\%) \\
English & 83597 (17.464\%) & 295530 (61.737\%) & 392949 (82.088\%) & 177345 (37.048\%) & 66364 (13.864\%) & 52955 (11.062\%) & 82376 (17.209\%) \\
Italian & 1327 (0.277\%) & 428 (0.089\%) & 1422 (0.297\%) & 284 (0.059\%) & 2937 (0.614\%) & 5047 (1.054\%) & 11123 (2.324\%) \\
Spanish & 939 (0.196\%) & 911 (0.19\%) & 1346 (0.281\%) & 662 (0.138\%) & 3223 (0.673\%) & 3674 (0.768\%) & 7363 (1.538\%) \\
Portuguese & 597 (0.125\%) & 534 (0.112\%) & 804 (0.168\%) & 417 (0.087\%) & 1727 (0.361\%) & 1846 (0.386\%) & 3130 (0.654\%) \\
German & 421 (0.088\%) & 263 (0.055\%) & 541 (0.113\%) & 168 (0.035\%) & 1232 (0.257\%) & 3627 (0.758\%) & 8517 (1.779\%) \\
Chinese & 164 (0.034\%) & 66 (0.014\%) & 128 (0.027\%) & 47 (0.01\%) & 1543 (0.322\%) & 965 (0.202\%) & 1404 (0.293\%) \\
Arabic & 134 (0.028\%) & 81 (0.017\%) & 147 (0.031\%) & 64 (0.013\%) & 2510 (0.524\%) & 1774 (0.371\%) & 3121 (0.652\%) \\
Russian & 97 (0.02\%) & 20 (0.004\%) & 81 (0.017\%) & 15 (0.003\%) & 426 (0.089\%) & 730 (0.152\%) & 1202 (0.251\%) \\
Romanian & 85 (0.018\%) & 18 (0.004\%) & 93 (0.019\%) & 12 (0.003\%) & 495 (0.103\%) & 827 (0.173\%) & 1148 (0.24\%) \\
Latin & 81 (0.017\%) & - & 57 (0.012\%) & 2 (0.0\%) & 65 (0.014\%) & 251 (0.052\%) & 254 (0.053\%) \\
Polish & 52 (0.011\%) & 23 (0.005\%) & 76 (0.016\%) & 15 (0.003\%) & 258 (0.054\%) & 427 (0.089\%) & 708 (0.148\%) \\
Greek & 38 (0.008\%) & 16 (0.003\%) & 41 (0.009\%) & 11 (0.002\%) & 436 (0.091\%) & 396 (0.083\%) & 723 (0.151\%) \\
Czech & 31 (0.006\%) & 23 (0.005\%) & 44 (0.009\%) & 12 (0.003\%) & 139 (0.029\%) & 206 (0.043\%) & 349 (0.073\%) \\
Catalan & 29 (0.006\%) & 16 (0.003\%) & 34 (0.007\%) & 10 (0.002\%) & 51 (0.011\%) & 145 (0.03\%) & 356 (0.074\%) \\
Japanese & 27 (0.006\%) & 14 (0.003\%) & 38 (0.008\%) & 16 (0.003\%) & 269 (0.056\%) & 178 (0.037\%) & 503 (0.105\%) \\
Hungarian & 25 (0.005\%) & 10 (0.002\%) & 23 (0.005\%) & 1 (0.0\%) & 101 (0.021\%) & 195 (0.041\%) & 259 (0.054\%) \\
Hebrew & 19 (0.004\%) & 1 (0.0\%) & 16 (0.003\%) & 1 (0.0\%) & 48 (0.01\%) & 143 (0.03\%) & 219 (0.046\%) \\
Breton & 17 (0.004\%) & 4 (0.001\%) & 16 (0.003\%) & 2 (0.0\%) & 47 (0.01\%) & 68 (0.014\%) & 56 (0.012\%) \\
Vietnamese & 17 (0.004\%) & 21 (0.004\%) & 24 (0.005\%) & 16 (0.003\%) & 450 (0.094\%) & 228 (0.048\%) & 290 (0.061\%) \\
Basque & 16 (0.003\%) & 11 (0.002\%) & 24 (0.005\%) & 9 (0.002\%) & 29 (0.006\%) & 38 (0.008\%) & 59 (0.012\%) \\
Bulgarian & 11 (0.002\%) & 4 (0.001\%) & 13 (0.003\%) & 4 (0.001\%) & 73 (0.015\%) & 139 (0.029\%) & 135 (0.028\%) \\
Persian & 10 (0.002\%) & 7 (0.001\%) & 8 (0.002\%) & 6 (0.001\%) & 140 (0.029\%) & 112 (0.023\%) & 257 (0.054\%) \\
Ukrainian & 10 (0.002\%) & 4 (0.001\%) & 7 (0.001\%) & 2 (0.0\%) & 75 (0.016\%) & 50 (0.01\%) & 76 (0.016\%) \\
\bottomrule
\end{tabular}
\end{table}
\end{landscape}

\section{Re-use of the dataset}

The dataset is designed to facilitate research on Ph.D. theses and academic interactions by providing as exhaustive and structured information as possible on doctoral theses defended in France between 1985 and 2025. Building upon the data initially made available by \textit{Thèses.fr}, we extend the original records in several ways. First, we introduce additional features derived from existing data, allowing for longitudinal and relational analyses. Second, we correct certain pre-existing features in order to reduce inconsistencies and improve data reliability for downstream research. Third, we enrich the dataset with new information collected from additional sources, including IdRef, \textit{Thèses en Ligne} (TEL), and SUDOC, thereby increasing the amount of available metadata and improving interoperability with other academic data infrastructures.

The resulting dataset is intended to support a wide range of research use cases related to doctoral education, academic careers, institutional collaboration, and the structure of research communities. One possible use case is the construction of graph-based representations of interactions between researchers, where nodes represent individuals and edges represent relationships such as thesis supervision or participation in thesis defense juries. Such graphs can be used to study the structure of academic communities, identify central actors, and detect sub-communities using graph clustering algorithms\footnote{For example, we provide here an example data visualization based on a subsample of the dataset: \url{https://acss-psl.github.io/DRMViz/}}. This type of analysis can provide insights into the organization of research fields, collaboration patterns, and the evolution of academic networks over time. Another use case consists of performing explanatory or statistical analyses based on the features included in the dataset. Because the dataset combines individual-level, thesis-level, and institutional-level variables, it enables the study of various aspects of doctoral education and academic participation. For example, in ongoing work, we use this dataset to conduct an explanatory analysis of the share of female researchers participating in thesis defense juries, using features derived from the dataset. More generally, the dataset can be used to study topics such as academic career trajectories, supervision practices, jury composition, institutional collaboration, internationalization of doctoral research, and the evolution of disciplines over time. The presence of multilingual metadata makes it also suitable for both French-focused and international or comparative research, as well as for text analysis and natural language processing tasks across multiple languages.

More broadly, the dataset was designed with reuse and extensibility in mind. The inclusion of persistent identifiers (such as TEL and SUDOC identifiers) allows future researchers to enrich the dataset with additional metadata, such as publication records, institutional rankings, disciplinary taxonomies, or full-text analysis of thesis manuscripts. As a result, the dataset should not be considered a static resource but rather a foundation that can be extended and adapted for new research questions related to higher education, research organization, and academic networks.

\printbibliography

\newpage
\appendix

\renewcommand\thefigure{\alph{figure}}
\renewcommand\theHfigure{\Alph{section}.\alph{figure}}
\setcounter{figure}{0}

\renewcommand\thetable{\alph{table}}
\renewcommand\theHtable{\Alph{section}.\alph{table}}
\setcounter{table}{0}

\renewcommand\thesection{\Alph{section}}

\titleformat{\section}
  {\normalfont\Large\bfseries}
  {Appendix \thesection}
  {1em}
  {}

\section{Percentages of missing data for selected features} \label{appendix:missing data}

\begin{figure}[htb]
    \centering
    \includegraphics[width=\linewidth]{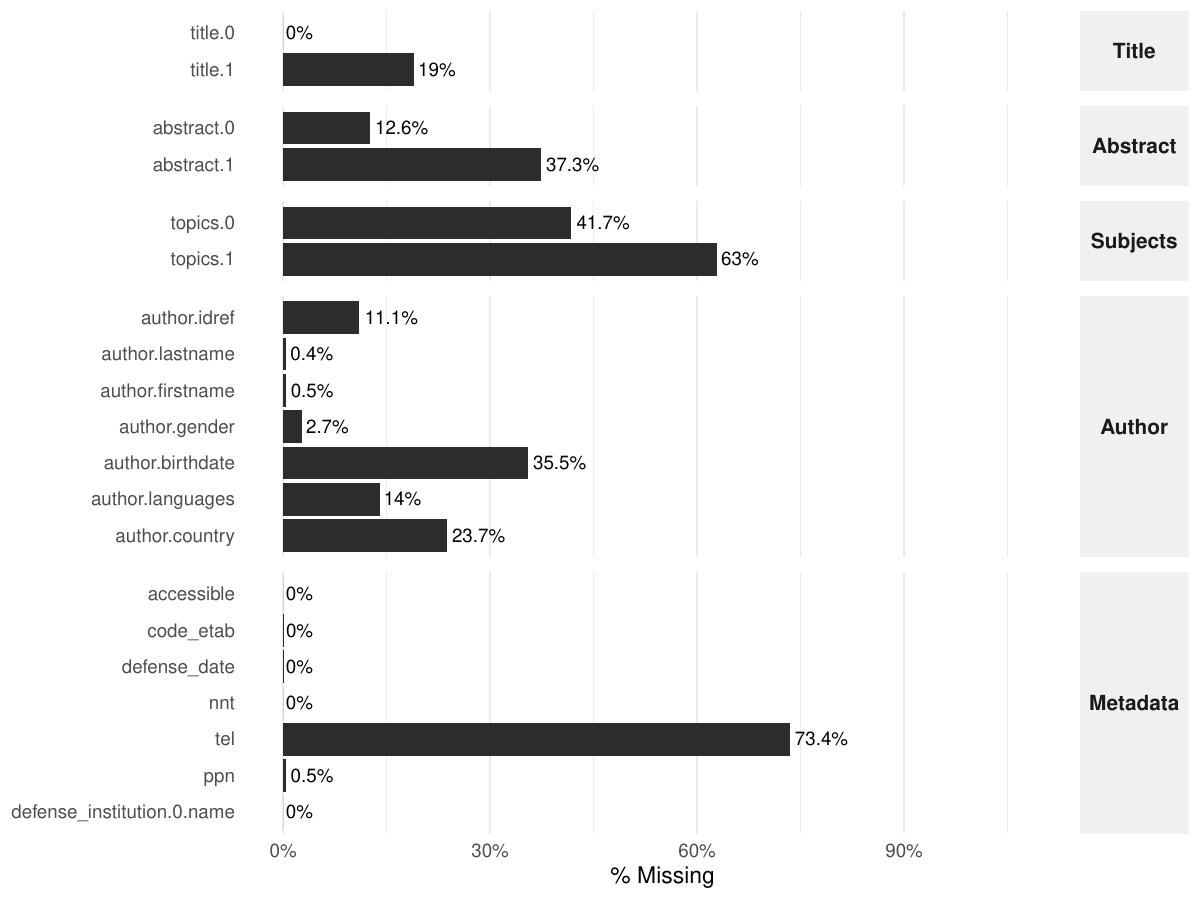}
    \caption{Data on theses missing}
    \label{fig:theses_missing}
\end{figure}
\begin{figure}[htb]
    \centering
    \includegraphics[width=\linewidth]{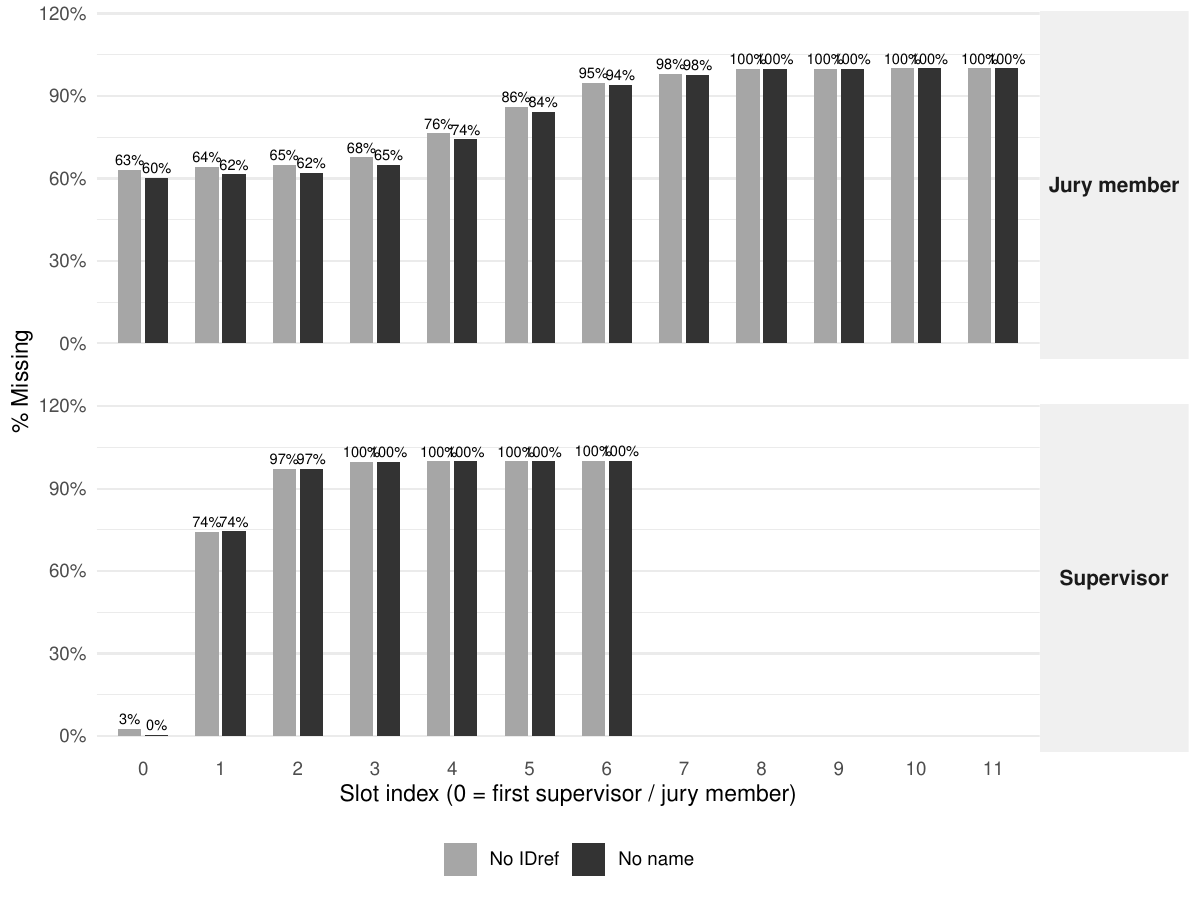}
    \caption{Data on supervisors and jury members missing}
    \label{fig:dir_jury_missing}
\end{figure}
\begin{figure}[htb]
    \centering
    \includegraphics[width=\linewidth]{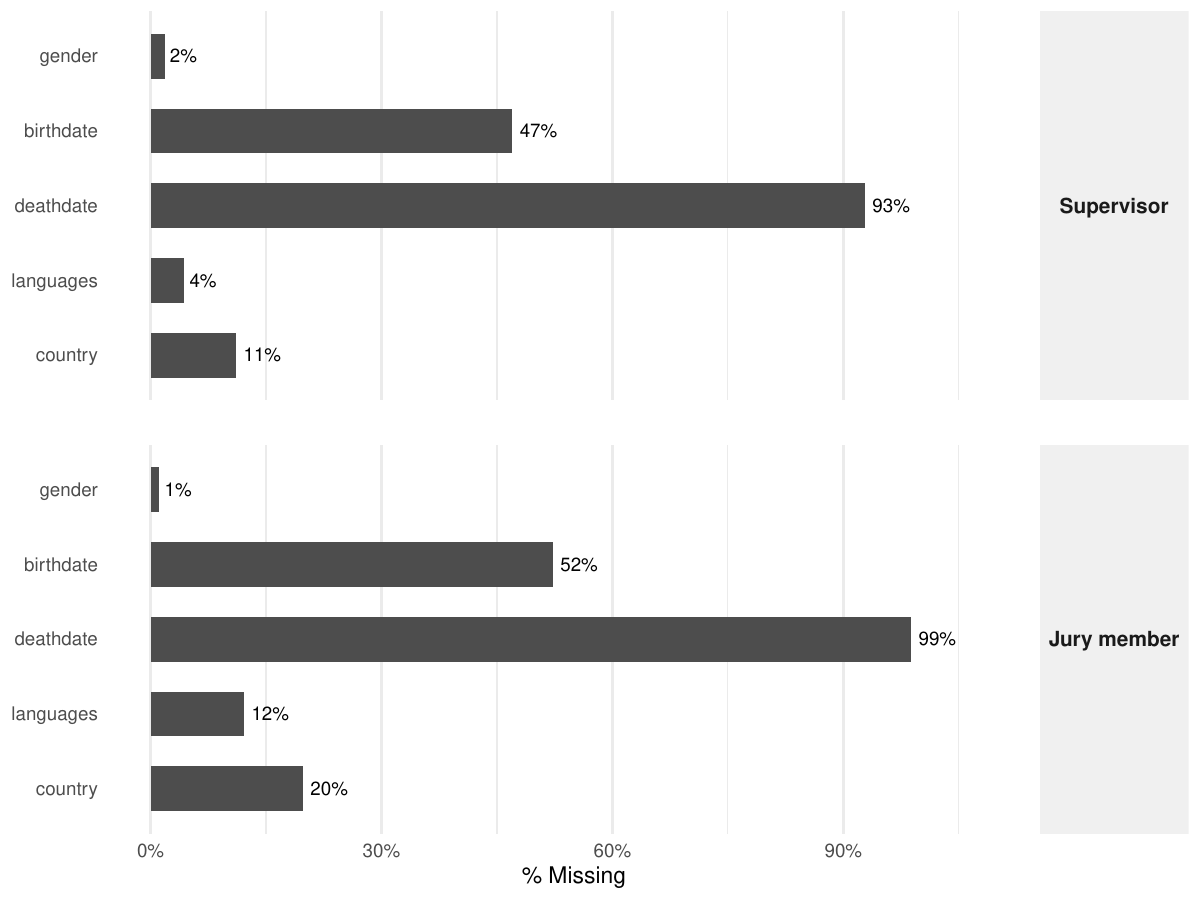}
    \caption{Enriched data from IdRef entries missing}
    \label{fig:extra_infos_missing}
\end{figure}
\begin{figure}[htb]
    \centering
    \includegraphics[width=\linewidth]{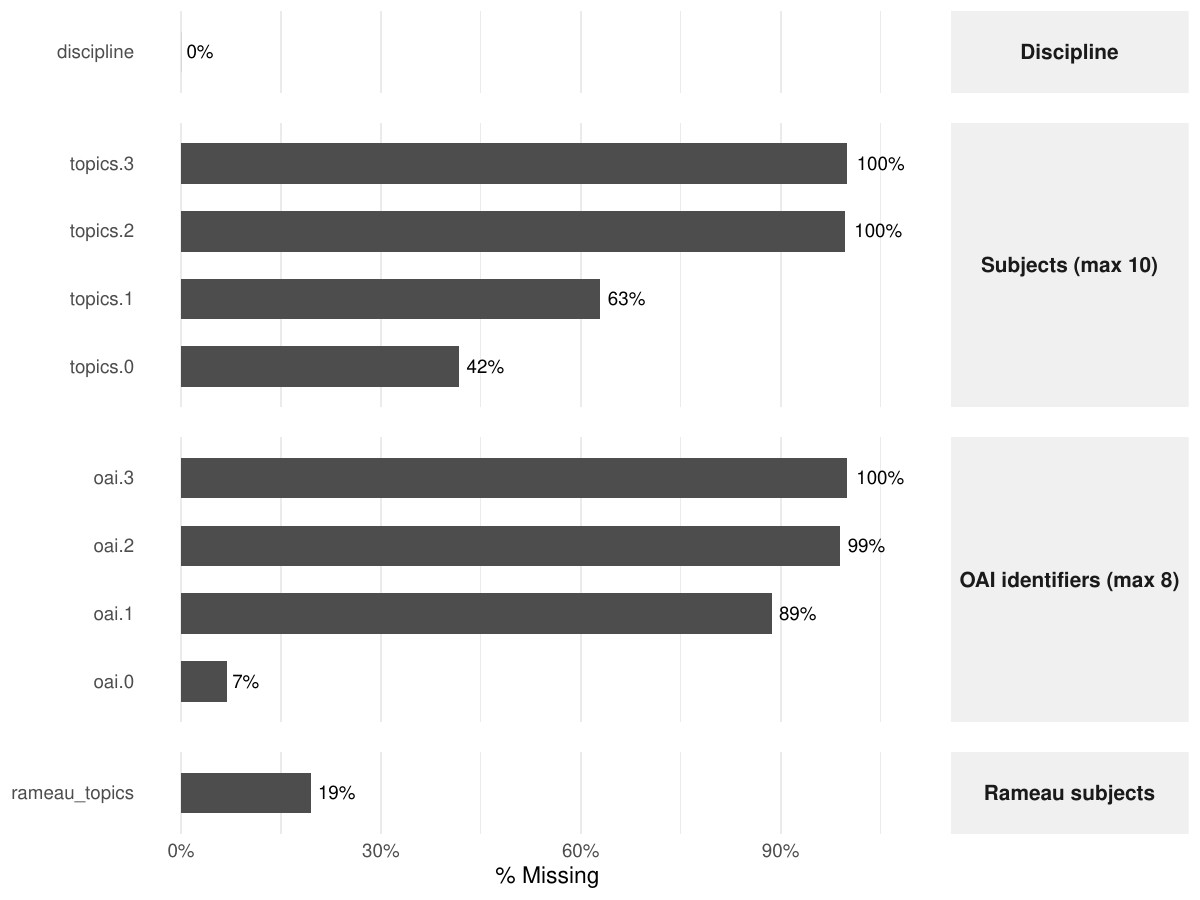}
    \caption{Subjects entries missing}
    \label{fig:subject_entries_missing}
\end{figure}

\FloatBarrier

\section{Variables codebook} \label{appendix:variables codebook}

\begin{longtable}{lllp{6.3cm}p{5cm}}
\caption{Overview of variable groups in the dataset}
\label{tab:variables}\\
\toprule
\textbf{Group} & \textbf{Type} & \textbf{N slots} & \textbf{Variables} & \textbf{Description} \\
\midrule
\endfirsthead

\multicolumn{4}{l}{\small\textit{(continued from previous page)}}\\
\toprule
\textbf{Group} & \textbf{Type} & \textbf{N slots} & \textbf{Variables} & \textbf{Description} \\
\midrule
\endhead

\midrule
\multicolumn{4}{r}{\small\textit{(continued on next page)}}\\
\endfoot

\bottomrule
\endlastfoot

\multicolumn{4}{l}{\textit{Thesis identifiers \& status}} \\*[0.3em]
& string & 1 & \texttt{nnt} & National thesis number (unique identifier) \\
& string & 1 & \texttt{tel} & \textit{Thèses en ligne} thesis number (unique identifier) \\
& string & 1 & \texttt{ppn} & SUDOC Pica Production Number (unique identifier) \\
& boolean & 1 & \texttt{accessible} & Whether the thesis is publicly accessible (\texttt{true}) or not (\texttt{false}) \\
& datetime & 1 & \texttt{embargo} & Embargo date if applicable. Not available for theses gathered using the API. \\
& string & 1 & \texttt{cas} & System used to publish the thesis record \\
& string & 1 & \texttt{source} & Data source \\
& boolean & 1 & \texttt{phd\_by\_publication} & \texttt{true} if the doctorate is a thesis by published works, \texttt{false} otherwise. Not available for theses gathered using the API. \\
& string & 1 & \texttt{code\_etab} & Institution code \\
& datetime & 1 & \texttt{defense\_date} & Thesis defense date \\
& boolean & 1 & \texttt{from\_api} & \texttt{true} if the thesis has been gathered from the Thèses.fr API, \texttt{false} if it is from the 2023 bulk export\\[0.5em]

\multicolumn{4}{l}{\textit{Author}} \\*[0.3em]
& string & 1 & \texttt{author.idref} & IdRef authority identifier \\
& string & 1 & \texttt{author.lastname}, \texttt{author.firstname} & Last and first name \\
& string & 1 & \texttt{author.gender} & Gender \\
& datetime & 1 & \texttt{author.birthdate}, \texttt{author.deathdate} & Birth and death date \\
& string & 1 & \texttt{author.languages} & Languages spoken \\
& string & 1 & \texttt{author.country} & Country of origin \\[0.5em]

\multicolumn{4}{l}{\textit{Supervisors}} \\*[0.3em]
& string & 0--6 & \texttt{supervisor.\{i\}.idref} & IdRef authority identifier \\
& string & 0--6 & \texttt{supervisor.\{i\}.lastname}, \texttt{supervisor.\{i\}.firstname} & Last and first name \\
& string & 0--6 & \texttt{supervisor.\{i\}.gender} & Gender \\
& datetime & 0--6 & \texttt{supervisor.\{i\}.birthdate}, \texttt{supervisor.\{i\}.deathdate} & Birth and death date \\
& string & 0--6 & \texttt{supervisor.\{i\}.languages} & Languages spoken \\
& string & 0--6 & \texttt{supervisor.\{i\}.country} & Country of origin \\
& number & 0--6 & \texttt{supervisor.\{i\}.centrality} & Number of occurrences in juries in the past 4 years \\
& number & 0--6 & \texttt{supervisor.\{i\}.age} & Age at time of the thesis defense \\
& number & 0--6 & \texttt{supervisor.\{i\}.yrs\_since\_phd} & Years since the person's thesis defense \\
& number & 0--6 & \texttt{supervisor.\{i\}.yrs\_since\_first\_jury} & Years since the person's first participation in a thesis jury in France \\
& number & 1 & \texttt{num\_supervisors} & Number of supervisors \\[0.5em]

\multicolumn{4}{l}{\textit{Jury}} \\*[0.3em]
& string & 0--11 & \texttt{jury\_member.\{i\}.idref} & IdRef authority identifier \\
& string & 0--11 & \texttt{jury\_member.\{i\}.lastname}, \texttt{jury\_member.\{i\}.firstname} & Last and first name \\
& string & 0--11 & \texttt{jury\_member.\{i\}.gender} & Gender \\
& datetime & 0--11 & \texttt{jury\_member.\{i\}.birthdate}, \texttt{jury\_member.\{i\}.deathdate} & Birth and death date \\
& string & 0--11 & \texttt{jury\_member.\{i\}.languages} & Languages spoken \\
& string & 0--11 & \texttt{jury\_member.\{i\}.country} & Country of origin \\
& number & 0--11 & \texttt{jury\_member.\{i\}.centrality} & Number of occurrences in juries in the past 4 years \\
& number & 0--11 & \texttt{jury\_member.\{i\}.age} & Age at time of the thesis defense \\
& number & 0--11 & \texttt{jury\_member.\{i\}.yrs\_since\_phd} & Years since the person's thesis defense \\
& number & 0--11 & \texttt{jury\_member.\{i\}.yrs\_since\_first\_jury} & Years since the person's first participation in a thesis jury in France \\
& number & 1 & \texttt{num\_jury\_members} & Number of jury members \\
& string & 1 & \texttt{jury\_president.idref} & Jury president IdRef identifier \\
& string & 1 & \texttt{jury\_president.lastname}, \texttt{jury\_president.firstname} & Jury president name \\
& string & 1 & \texttt{jury\_president.gender} & Jury president gender \\
& datetime & 1 & \texttt{jury\_president.birthdate}, \texttt{jury\_president.deathdate} & Jury president birth and death date \\
& string & 1 & \texttt{jury\_president.languages} & Jury president languages \\
& string & 1 & \texttt{jury\_president.country} & Jury president country \\
& string & 0--5 & \texttt{rapporteur.\{i\}.idref} & Rapporteur IdRef identifier \\
& string & 0--5 & \texttt{rapporteur.\{i\}.lastname}, \texttt{rapporteur.\{i\}.firstname} & Rapporteur name \\
& string & 0--5 & \texttt{rapporteur.\{i\}.gender} & Rapporteur gender \\
& datetime & 0--5 & \texttt{rapporteur.\{i\}.birthdate}, \texttt{rapporteur.\{i\}.deathdate} & Rapporteur birth and death date \\
& string & 0--5 & \texttt{rapporteur.\{i\}.languages} & Rapporteur languages \\
& string & 0--5 & \texttt{rapporteur.\{i\}.country} & Rapporteur country \\
& number & 1 & \texttt{num\_rapporteur} & Number of thesis rapporteurs \\[0.5em]

\multicolumn{4}{l}{\textit{Institutional affiliations}} \\*[0.3em]
& string & 0--4 & \texttt{defense\_institution.\{i\}.idref} & Defense institution IdRef identifier \\
& string & 0--4 & \texttt{defense\_institution.\{i\}.name} & Defense institution name \\
& string & 0--1 & \texttt{doctoral\_school.\{i\}.idref} & Doctoral school IdRef identifier \\
& string & 0--1 & \texttt{doctoral\_school.\{i\}.name} & Doctoral school name \\
& string & 0--7 & \texttt{research\_partner.\{i\}.idref} & Research partner IdRef identifier \\
& string & 0--7 & \texttt{research\_partner.\{i\}.name} & Research partner name \\
& string & 0--7 & \texttt{research\_partner.\{i\}.type} & Research partner type \\
& number & 1 & \texttt{num\_research\_partners} & Number of research partners \\[0.5em]

\multicolumn{4}{l}{\textit{Content \& topics}} \\*[0.3em]
& string & 0--5 & \texttt{title.\{i\}}, \texttt{title.\{i\}.language} & Thesis title and its language \\
& string & 0--7 & \texttt{abstract.\{i\}}, \texttt{abstract.\{i\}.language} & Abstract and its language \\
& string & 0--9 & \texttt{topics.\{i\}}, \texttt{topics.\{i\}.language} & Keywords and their language \\
& string & 1 & \texttt{discipline} & Discipline field \\
& string & 1 & \texttt{rameau\_topics} & Rameau \cite{BNFRameau2017} topics headings (separated with \texttt{||}) \\
& string & 0--3 & \texttt{language.\{i\}} & Thesis language(s) \\
& string & 0--7 & \texttt{oai.\{i\}} & Dewey Decimal Classes. Not available for theses gathered using the API. \\

\end{longtable}

\end{document}